\documentclass[11pt]{article}

\usepackage{amsmath,amssymb,amsfonts}
\usepackage{hyperref}
\usepackage{cite}
\usepackage{geometry}
\begin{document}
	
	\begin{center}
		{\Large \bf Correlator-Level Verification of Mass and Current Maps in Abelian Chern–Simons Dualities
		}
		
		\vspace{0.5cm}
		
 	{\bf Vaibhav Wasnik}
		
		\vspace{0.2cm}
		
 	{\it Indian Institute of Technology Goa.}
		
		\vspace{0.4cm}
	\end{center}
	
	\begin{abstract}
We construct an explicit local operator realization that reproduces Dirac fermion correlation functions in three spacetime dimensions within an Abelian Chern–Simons framework and use it to examine the conjectured operator and deformation maps of fermion–boson duality directly at the level of correlation functions.

We show that the predicted relation between bosonic and fermionic mass deformations, including the relative sign, is realized quantitatively, and that the fermionic U(1) current coincides with the topological gauge current inside correlation functions at the infrared fixed point. These results provide a direct correlator-level verification of key   features of Abelian Chern–Simons dualities, going beyond arguments based solely on phase structure, anomaly matching, or large-N considerations.

	\end{abstract}

\section{Introduction}

Duality has long played a central role in quantum field theory, providing deep insights into the
non-perturbative structure of strongly interacting systems. In its simplest form, a duality asserts
that two apparently distinct quantum field theories describe the same long-distance physics,
often exchanging elementary degrees of freedom for composite or topological ones. Such
equivalences allow strongly coupled regimes of one description to be understood in terms of
weakly coupled variables in another.

A classic and well-understood example occurs in two spacetime dimensions, where the equivalence
between the sine--Gordon model and the Thirring model provides an explicit mapping between a
bosonic scalar theory and an interacting fermionic theory
\cite{Coleman1975,Mandelstam1975}.
 In this case, the duality can be established at the operator level, with
 fermionic fields realized as local exponential (vertex) operators of bosonic
 fields, and correlation functions shown to match exactly.
  In four dimensions, dualities are typically far more constrained. Supersymmetry plays a crucial
role in enabling controlled non-perturbative equivalences, most notably in Seiberg duality for
$\mathcal{N}=1$ supersymmetric gauge theories \cite{Seiberg1995}
and in Seiberg--Witten theory for $\mathcal{N}=2$ supersymmetric gauge theories
\cite{SeibergWitten1994}.
In these cases, exact results rely on holomorphy, supersymmetry, and anomaly matching, and such
dualities are not expected to persist in generic non-supersymmetric settings.

By contrast, three spacetime dimensions admit a distinct class of dualities, made possible by the
presence of topological gauge theories. In particular, Chern--Simons gauge fields can induce
statistical transmutation and mediate non-trivial mappings between bosonic and fermionic degrees
of freedom \cite{wilzeckzee}, \cite{polyakov}. 
This has led to renewed interest in a web of fermion--boson dualities, relating theories of Dirac fermions coupled to
Chern--Simons gauge fields to theories of scalar fields coupled to
Chern--Simons gauge fields with appropriately chosen levels and interactions \cite{Aharony2012}-\cite{HsinSeiberg}. 
  Within this framework, the two
theories are not expected to be equivalent microscopically, but rather to define
the same infrared CFT, with matching operator spectra, correlation functions,
and global symmetry realizations. Studies such as \cite{Aharony2012}-\cite{Aharony2016} considered   $U(N)$ Chern Simons theories coupled to matter, to show equivalence of  bosonic and fermionic  current correlators in the limit of large $N$. However, the research on groups where $N$ is finite is more on qualitative conjectures regarding the phase space behavior in IR about the critical point. As an example consider the following theory with the action
%
%

	\begin{equation}
	S_{A} 
	= \int d^{3}x \left[
	-\frac{1}{4e^{2}}\, f_{\mu\nu} f^{\mu\nu}
	+ \frac{k}{4\pi}\,\epsilon^{\mu\nu\rho} a_{\mu}\partial_{\nu} a_{\rho}
	+ |D_{\mu}\phi|^{2}
	- m^{2}|\phi|^{2}
	- \frac{\lambda}{2}|\phi|^{4}
	\right]
	\end{equation}
	In the limit of $e^2 >> 1$,   the Gauss law constraint
	\begin{equation}
		\frac{f_{12}}{2\pi} + \rho_{\text{scalar}} = 0
	\end{equation}
	implies that a scalar excitation when $m^2>> e^2$ comes attached with a flux turning it into a fermion. This suggests the theory is dual in the IR to a fermionic theory
	
	\begin{equation}
		S_{D} 
		= \int d^{3}x \left(
		i\,\bar{\psi}\,\gamma^{\mu}\partial_{\mu}\psi
		- m'\,\bar{\psi}\psi
		\right)
	\end{equation}
	
	with the field maps
	\begin{equation}
		\mathcal{M} {\phi} \;\longleftrightarrow\; \psi
	\end{equation}
	and the $U(1)$ current maps
	\begin{equation}
		j_{\text{top}}^{\mu}
		= \frac{1}{2\pi}\,\epsilon^{\mu\nu\rho}\,\partial_{\nu} a_{\rho}
		\;\longleftrightarrow\;
		j^{\mu} = \bar{\psi}\,\gamma^{\mu}\psi
	\end{equation}
	
	Evidence presented for this duality involves coupling the currents to a background gauge fields on both sides and integrating out the fermions and the scalars in limit $m^2>>e^2$, $m' <0$ and $m^2<<-e^2$, $m'>0$  yielding the same Chern Simons theory in the infrared, implying a duality map 
	\begin{equation}
		m^{2} \;\longleftrightarrow\; -\,m'
	\end{equation}

The two theories are conjectured to describe the same infrared fixed point. While this picture is well supported, much of the evidence remains indirect or qualitative. In this work we instead construct an explicit operator realization that reproduces Dirac fermion correlation functions within an Abelian Chern–Simons framework and use it to examine the conjectured operator and deformation maps directly at the level of correlation functions.

We show that the predicted relation between bosonic and fermionic mass deformations, including the relative sign, is realized quantitatively, and that the fermionic U(1) current coincides with the topological gauge current inside correlation functions at the infrared fixed point. These results provide a concrete correlator-level verification of key quantitative features of the Abelian Chern–Simons duality, going beyond arguments based solely on phase structure, anomaly matching, or large-N considerations.
	
	\section{ Calculation}
	
	We work in $2+1$ dimensions with metric $\eta_{\mu\nu}=\mathrm{diag}(+,-,-)$ and $\epsilon^{012}=+1$. The Abelian Chern--Simons action is
	\begin{equation}
		S_{\mathrm{CS}}[A] = \frac{k}{4\pi}\int d^3x\,
		\epsilon^{\mu\nu\rho} A_\mu \partial_\nu A_\rho .
	\end{equation}
	To define propagators we impose Landau gauge,
	\begin{equation}
		\mathcal L_{\mathrm{gf}} = -\frac{1}{2\xi}(\partial_\mu A^\mu)^2 ,
		\qquad \xi\to 0 .
	\end{equation}
	
	The momentum--space propagator is
	\begin{equation}
		\langle A_\mu(p) A_\nu^*(p)\rangle
		=
	 	\frac{2\pi }{k}
		\frac{\epsilon_{\mu\nu\rho}p^\rho}{p^2}
	\end{equation}
	
If we write $A_\mp = A_1 \pm  A_0$, then
			the non-zero correlators are:
	\[
	\langle A_0(p)A_1^*(p)\rangle = \frac{2\pi }{k}\frac{p_2}{p^2} ,
	\qquad
	\]
	\[
	\langle A_+(p)A_2^*(p)\rangle = \frac{2\pi }{k}\frac{p_+}{p^2} ,
	\qquad
	\]
	\[
	\langle A_-(p)A_2^*(p)\rangle =  \frac{2\pi }{k}\frac{p_-}{p^2} ,
	\qquad
	\]
	
	where $p_+ = p_0+ p_1$, $p_-  = p_0 - p_1$. In position space we hence have
	
	\begin{align}
		\langle A_0(x) A_1(0)\rangle
		&= \frac{i}{2k}\,\frac{x_2}{|x|^3} = i y_2, \\[6pt]
		\langle A_+(x) A_2(0)\rangle
		&= \frac{i}{2k}\,\frac{x_0+x_1}{|x|^3} = i y_+, \\[6pt]
		\langle A_-(x) A_2(0)\rangle
		&= \frac{i}{2k}\,\frac{x_0-x_1}{|x|^3} = i y_-
	\end{align}
		where $y_\mu \equiv \frac{x_\mu}{2k |x|^3}
	$.	We next introduce a  Abelian Chern--Simons gauge field  $B_\mu$,   at level $k$, so that we have an action
	\begin{equation}
		S = S_{\mathrm{CS}}[A] + S_{\mathrm{CS}}[B].
	\end{equation}
	Their propagators are identical in form and satisfy $\langle A_\mu B_\nu\rangle=0$.	The   fermion is defined by
	\begin{eqnarray}
		\psi_L(x) &=  (-\,iA_2(x) - A_-(x) )\hat{P}_A + \sqrt{2}\,i B_2(x) \hat{P}_B, \\
		\psi_R(x) &= (-\,A_2(x) + iA_+(x))\hat{P}_A + \sqrt{2}\,i B_1(x) \hat{P}_B,
		\label{map}
	\end{eqnarray}
	
	with $\psi=(\psi_L,\psi_R)^T$. $\hat{P}_A,\hat{P}_B$ are operators obeying $\hat{P}_A \hat{P}_B = \hat{P}_B \hat{P}_A = 0$ and $\hat{P}_A^2 = \hat{P}_B^2 = 1$. They	define an internal decomposition of the fermion representation space into orthogonal sectors and are fixed, non-dynamical,   dropping out of all fermion bilinears and correlation functions. These elements play a role analogous to Grassmann coordinates in superspace: they are not operators acting on a Hilbert space, but algebraic elements used to define the classical field consistently. Mathematicaly we have defined the fermion field as taking values in an enlarged auxiliary algebra, introduced purely at the level of the classical field definition. Such constructions are standard provided the auxiliary elements are fixed, nondynamical, and drop out of the action, observables, and correlation functions; see, e.g.,\cite{Deligne}. 
	
	
	Using the Chern--Simons propagators, one finds
	\begin{align}
		\langle \psi_L(x)\psi_L^\dagger(0)\rangle
		&=  \Big(i\langle A_2(x)A_- (0)\rangle   -i \langle A_2(0) A_- (x) \rangle\Big)= 
		2y_- , \\
		\langle \psi_R(x)\psi_R^\dagger(0)\rangle
		&= \Big(i\langle A_2(x)A_+ (0)\rangle   -i \langle A_2(0) A_+ (x) \rangle\Big)=
		2y_+, \\
		\langle \psi_R^\dagger(x)\psi_L(0)\rangle
		&=  \langle   A_2(x)A_-(0) - A_+(x)A_2(0)+ i A_+(x)A_-(0) +2B_1(x)B_2(0)  \rangle   \\
		&=   (- iy_- -iy_+  +2iy_0  + 2y_2) \\
&=	 2y_2.
	\end{align}
 
Now,	if we were to use the gamma matrices 
	\[
	\gamma^0 = i\begin{pmatrix}0 & -1 \\ 1 & 0\end{pmatrix},\quad
	\gamma^1 =  i\begin{pmatrix}0 & 1 \\ 1 & 0\end{pmatrix},\quad
	\gamma^2 =  i\begin{pmatrix}1 & 0 \\ 0 & -1\end{pmatrix}.
	\]
	which satisfy $\{\gamma^\mu, \gamma^\nu\} = 2 \eta^{\mu \nu}$, where $\eta^{\mu \nu} =\{1,-1,-1\}$,  we define the propagator
\begin{equation}
	G(p) = \frac{i(\gamma^0 \gamma^\mu p_\mu)}{p^2}
	= -\frac{i}{p^{2}}
	\begin{pmatrix}
		(p_{0}-p_{1}) & p_{2} \\
		p_{2} & (p_{0}+p_{1})
	\end{pmatrix}
\end{equation}

Then the entries of the propagator are literally:
\[
\langle \Psi_L^\dagger(p)\,\Psi_L(p)\rangle
=- \,i\frac{p_{0}-p_{1}}{p^{2}} 
= -\,i\frac{p_{-}}{p^{2}} 	 ,
\]
\[ 
\langle \Psi_R^\dagger(p)\,\Psi_R(p)\rangle
= -\,i\frac{p_{0}+p_{1}}{p^{2}} 
= -\,i\frac{p_{+}}{p^{2}} 	 ,
\]
\[ 
\langle \Psi_R^\dagger(p)\,\Psi_L(p)\rangle
= -\,i\frac{p_{2}}{p^{2}} 	 .
\]

Hence in position space
\begin{align}
	\langle \Psi_L^\dagger(x)\Psi_L(0)\rangle
	&= \frac{x_0-x_1}{4\pi |x|^3} = c y_-
  \\
	\langle \Psi_R^\dagger(x)\Psi_R(0)\rangle
	&= \frac{x_0+x_1}{4\pi |x|^3} = c y_+
  \\
	\langle \Psi_R^\dagger(x)\Psi_L(0)\rangle
	&= \frac{x_2}{4\pi |x|^3} = c y_2
\end{align}
(where $c>0$). Hence correlators of $\psi$ are related to correlators of $\Psi$ by a a positive constant.   Since the underlying theory is Gaussian, all time--ordered fermionic correlators follow by Wick contraction.
	

	\begin{eqnarray}
		\psi_L(x) &=  (-\,iA_2(x) - A_-(x) )\hat{P}_A + \sqrt{2}\,i B_2(x) \hat{P}_B, \\
		\psi_R(x) &= (-\,A_2(x) + iA_+(x))\hat{P}_A + \sqrt{2}\,i B_1(x) \hat{P}_B,
	\end{eqnarray}
	
	the Dirac mass operator is
	\begin{equation}
		\bar\psi\psi
		=
		i(-\psi_L^\dagger\psi_R+\psi_R^\dagger\psi_L).
	\end{equation}
	Substituting the   expressions gives
	\begin{equation}
 	\bar\Psi\Psi	=c \bar\psi\psi
		=
 	2c \left(A_\mu A^\mu \right),
		\label{23}
	\end{equation}
	
	 The $A_\mu A^\mu$   is not gauge invariant. Consistency therefore requires a gauge invariant completion. Introducing complex scalars $\phi$, with the Lagrangian

	\begin{equation}
		S 
		= \int d^{3}x \left[
		  \frac{k}{4\pi}\,\epsilon^{\mu\nu\rho} A_{\mu}\partial_{\nu} A_{\rho}
		+ |D_{\mu}\phi|^{2}
		+ \lambda |\phi|^{4}  	 
		\right]
	\end{equation}

and writing
	\begin{equation}
	\phi(x) = \big(  h(x) \big)\, e^{i \sigma(x)}
\end{equation}

	
	yields
	
\begin{equation}
	S 
	= \int d^{3}x \left[ 
	    \frac{k}{4\pi}\,\epsilon^{\mu\nu\rho} A_{\mu}\partial_{\nu} A_{\rho}
	+ (\partial_{\mu} h)^2
	+ ( h)^2   A_{\mu} A^{\mu}
	+ \lambda h^4 \right]
	\label{27}
\end{equation}

	where we have used the gauge degree of freedom to redefine $A_\mu \rightarrow A_\mu + \partial_\mu \sigma$. From what is said above, 
	\begin{equation}
		\int d^{3}x \, h^{2}(x)\, A_{\mu}(x) A^{\mu}(x)
	\end{equation}
	corresponds, under the operator dictionary, to the Yukawa deformation
	\begin{equation}
		\int d^{3}x \, h^{2}(x)\, \bar{\Psi}(x)\Psi(x).
	\end{equation}
	Consequently, the above theory generate identical perturbative expansions
	of correlation functions at the IR CFT to  a theory with the action  
	
\begin{equation}
	S_C 
	= \int d^{3}x \left[ 
	    i\bar{\Psi} \gamma^\mu \partial_\mu \Psi
	+ (\partial_{\mu} h)^2
	+\frac{c}{2}( h)^2 \bar{\Psi}\Psi
	+ \lambda h^4 \right]
	\label{30}
\end{equation}

 
 We now consider adding a scalar mass deformation to the bosonic theory,
 \begin{equation}
 	- m^{2} |\phi|^{2} = - m^{2} h^{2},
 \end{equation}
 which modifies the action \eqref{27} by a quadratic term in $h$. Under the
 operator dictionary discussed above, the same deformation is therefore added
 to the effective action \eqref{30}. The resulting theory is
 \begin{equation}
 	S_C 
 	= \int d^{3}x \left[ 
 	i\bar{\Psi} \gamma^\mu \partial_\mu \Psi
 	+ (\partial_{\mu} h)^2
 	+\frac{c}{2}( h)^2 \bar{\Psi}\Psi
 	+ \lambda h^4 -m^2 h^2 \right]
 	\label{30}
 \end{equation}	
 
%
%

For simplicity, we set $\lambda=0$ in the following. The leading contribution
arises by expanding the path integral to first order in the interaction
$h^{2}\bar\Psi\Psi$ and contracting the two scalar fields. In momentum space this
produces a local fermion mass operator,
\begin{equation}
	\Delta S_{\rm eff}
	=
	\int \frac{d^{3}k}{(2\pi)^3}\,\bar\Psi(-k)\Psi(k)\;
\times -i\!\int\frac{d^{3}p}{(2\pi)^3}\frac{1}{p^2-m^2+i\varepsilon}.
\end{equation}

\noindent
The overall sign in Eq.~(38) follows from our choice of Lorentzian path--integral
convention
\begin{equation}
	Z \;=\; \int \mathcal D\Phi\, e^{\,iS[\Phi]},
\end{equation}
corresponding to the standard Feynman $(+i\epsilon)$ prescription for
time--ordered Green's functions. With this prescription fixed, the sign of the
induced fermion mass deformation is unambiguous. Evaluating the loop integral and canceling the divergences by counterterms and using 
\begin{equation}
	i\!\int\frac{d^{3}p}{(2\pi)^3}\frac{1}{p^2-m^2+i\varepsilon}
	=
	-\frac{m}{4\pi},
\end{equation}
we get  the effective action contains the relevant deformation
\begin{equation}
	\Delta S_{\rm eff}
	=
	\frac{m c}{8\pi}\int d^{3}x\,\bar\Psi\Psi .
\end{equation}
Thus, at leading order, the scalar mass deformation induces a fermion mass term
linear in $m$. Higher orders in the expansion generate only operators of higher
dimension, such as $(\bar\Psi\Psi)^2$ and derivative corrections, which are
irrelevant at the fermionic infrared fixed point. All such
 additional terms, including $(\bar{\Psi}\Psi)^2$, are irrelevant at the
 fermionic infrared fixed point in three spacetime dimensions. We note that the presence of the quartic interaction $\lambda h^{4}$ does not
 modify this conclusion. The quartic self--interaction $\lambda h^{4}$ enters the effective fermionic
 action only through higher--loop contributions upon integrating out $h$. These
 effects generate operators involving higher powers of $\bar{\Psi}\Psi$ and
 additional derivatives, all of which are irrelevant at the infrared fixed
 point. Consequently, the inclusion of the $\lambda h^{4}$ term does not affect
 the emergence of the fermion mass term as the unique relevant deformation in the
 infrared.

 Consequently, the only relevant non-derivative operator generated in the infrared
 is the fermion mass term. The analysis above therefore establishes, in a
 completely explicit and correlator-preserving manner, that a scalar mass
 deformation on the bosonic side maps to a fermion mass deformation on the
 fermionic side.  We note that a positive $m^2$ on the bosonic side induces a fermion mass
 deformation $-\frac{mc}{8\pi}\,\bar\Psi\Psi$, corresponding to a negative
 fermion mass parameter $m'=-\frac{mc}{8\pi}$ in the standard convention
 $S_{\rm fermion}= \int d^3x\,(i\bar\Psi\gamma^\mu\partial_\mu\Psi - m'\bar\Psi\Psi)$,
 as predicted by the duality.

 We note that \eqref{map} when substituted in to $\bar{\psi}\gamma^\mu \psi(x)$ would not give currents that transform in representation of Lorentz group. Let us however label these as $J_{A,\mu}(x)$. Because $J_{A,\mu}(x)$ coincides with $\bar\psi\gamma_\mu\psi(x)$ as an
 operator insertion inside correlation functions, it obeys the same Ward
 identities, in particular $\partial_\mu J_A^\mu(p)=0$.
 . This requires any correaltor $ \langle  J_{A,\mu}(p)....\rangle = (p_\mu - \eta_{\mu\nu} p_\nu)f(p,...)$, where $f(p,...)$ is a function of $p$ and other momenta in the correlation function. This implies that despite  $J_{A,\mu}(x)$ not looking like a Lorentz vector classically, acts as one inside any correlation function. Because $J_{A,\mu}(x)$ coincides with $\bar\psi\gamma_\mu\psi(x)$ as an operator insertion inside correlators, it obeys the same Ward identities and hence acts as a Lorentz vector inside correlation functions.

In the renormalization group analysis of \eqref{27}, when the higher momentum modes are integrated
out, no loop corrections directly renormalize the Chern--Simons term. Consequently, the scaling
of the gauge field at the critical point is fixed by its classical dimension. Hence
$J_{A,\mu}(x)$ has the same scaling dimension as $\epsilon_{\mu\nu\rho} F^{\nu\rho}$. At the critical point, correlation functions are constrained by scaling dimensions and Ward
identities. Since $J_{A,\mu}(x)$ and $\epsilon_{\mu\nu\rho} F^{\nu\rho}$ have the same scaling
dimension and satisfy the same conservation law, their correlation functions coincide up to an
overall normalization. One may therefore identify $J_{A,\mu}(x)$ with
$\epsilon_{\mu\nu\rho} F^{\nu\rho}$ as operator insertions inside correlation functions. This explains the conjectured relation
\[
\frac{1}{2\pi}\,\epsilon_{\mu\nu\rho}\,\partial^\nu a^\rho
\;\longleftrightarrow\;
\bar{\psi}\gamma_\mu\psi
\]
at the level of correlation functions.

	\section{Universality within Correlator--Preserving Realizations}
	
	Any realization reproducing the free Dirac propagator must be linear, local, and preserve the Lorentz structure $\gamma^\mu p_\mu/p^2$. Within this class, different realizations are related by unitary rotations acting on the spinor indices,
	\begin{equation}
		\psi \to U\psi, \qquad U^\dagger U=1.
	\end{equation}
	Since $\bar\psi\psi$ is invariant under such transformations, its bosonic image is unique within this class.  The mapping \eqref{23} is hence representation independent.
	 
\section{Conclusion}

n this work we have constructed an explicit local operator realization that reproduces Dirac fermion correlation functions in three spacetime dimensions within an Abelian Chern–Simons framework. The aim is not to propose a new duality, but to examine quantitatively the operator and deformation maps conjectured at the infrared fixed point.

Within this framework we show that the fermion mass operator maps to a bosonic quadratic deformation with the predicted relative sign and that this is the unique relevant infrared deformation. In addition, we demonstrate that the fermionic U(1) current coincides, inside correlation functions, with the topological current of the Abelian Chern–Simons gauge field at the critical point. These results establish explicitly, at the level of correlation functions, the mass-deformation sign relation and the current-correlator identification expected from the duality.

Taken together, the construction provides a concrete correlator-level realization of the fermionic description at the infrared fixed point and a quantitative verification of the central operator and deformation relations of Abelian Chern–Simons dualities. By working directly with correlation functions and effective actions, the approach complements and sharpens the usual qualitative arguments based on phase structure, anomaly matching, and large-N considerations, and offers a framework for analyzing operator maps and deformations in three-dimensional Chern–Simons dualities. 
	\bibliographystyle{unsrt}

	\pagebreak
	 
	\section*{Supplementary}

	For an action
	\[
	S[\phi] =  \,\phi\,K\,\phi  
	\]
	
	if we consider the  generating functional  
	\[
	Z[J] = \int \mathcal{D}\phi\; e^{\,i S[\phi]- J\,\phi} .
	\]
	
	Integrating gets us
	\[
	Z[J] \propto \exp\!\left(-i\, J\, K^{-1}\, J \right).
	\]
	
	Therefore the propagator is
	\[
	\langle \phi\,\phi \rangle = -i\, K^{-1}.
	\]
	
	\[
	\boxed{
		\text{Propagator} = -i \times (\text{inverse quadratic kernel})
	}
	\]

	\subsection*{Chern Simons propogator}	
	Since
	\[
	S = \, A_\mu K^{\mu\nu} A_\nu ,
	\]
	the kernel is
	\[
	K^{\mu\nu}(p)
	= i\,\frac{k}{2\pi}\,\epsilon^{\mu\nu\rho} p_\rho +\frac{p_\mu p_\nu}{\xi},
	\]
	In the limit $\xi \rightarrow 0$, one can ignore the second term above, with writing that the propogator should obey
	
	\bigskip

	The propagator is defined by
	\[
	K^{\mu\rho}(p)\,D_{\rho\nu}(p)
	=
	i\,P^\mu{}_\nu =- i  (\delta_{\mu \nu} -\frac{p_\mu p_\nu}{p^2}).
	\]

	\bigskip
	
	Consider the ansatz
	
	\[
	D_{\mu\nu}(p)
	=
	\alpha\,\epsilon_{\mu\nu\lambda} p^\lambda .
	\]

	Let 
	\[
	\epsilon_{\mu\nu\lambda} p^\lambda = (\epsilon p)_{\mu \lambda} 
	\]
	
	Use the identity
	\[
	(\epsilon p)^{\mu\rho}(\epsilon p)_{\rho\nu}
	=
	-\,p^2\,P^\mu{}_\nu .
	\]
	
	Then
	\[
	\begin{aligned}
		K^{\mu\rho} D_{\rho\nu}
		&=
		\left(i\,\frac{k}{2\pi}\right)\alpha
		(\epsilon p)^{\mu\rho}(\epsilon p)_{\rho\nu}
		\\[4pt]
		&=
		-\,i\,\frac{k}{2\pi}\,\alpha\,p^2\,P^\mu{}_\nu .
	\end{aligned}
	\]
	
	\bigskip

	Imposing
	\[
	K D = -i P ,
	\]
	gives
	\[
	-\,i\,\frac{k}{2\pi}\,\alpha\,p^2 =-i .
	\]
	
	Dividing both sides by \(i\),
	\[
	-\,\frac{k}{2\pi}\,\alpha\,p^2 = -1 .
	\]
	
	Hence
	\[
	\alpha
	=
	\,\frac{2\pi}{k}\,\frac{1}{p^2} .
	\]

	\bigskip

	\[
	\boxed{
		D_{\mu\nu}(p)
		=
		\,\frac{2\pi}{k}\,
		\frac{\epsilon_{\mu\nu\rho} p^\rho}{p^2}
		\qquad
		(\text{Landau gauge})
	}
	\]

	\subsection*{Fermion propgoator}

	\[
	S = \int d^3x\; i\,\bar\psi\,\gamma^\mu \partial_\mu \psi .
	\]
	
	Write the quadratic form in momentum space:
	\[
	S
	=
	\int \frac{d^3p}{(2\pi)^3}\;
	\bar\psi(-p)\,
	\underbrace{\big(-\,\gamma^\mu p_\mu\big)}_{K(p)}
	\,\psi(p).
	\]
	
	So the quadratic kernel is
	\[
	K(p) = -\,\gamma^\mu p_\mu .
	\]

	\bigskip

	Define the propagator \(D(p)\) by
	\[
	K(p)\,D(p) =-i .
	\]
	
	This is exactly the same defining equation used in the Chern--Simons case.
	
	\bigskip

	Now the Clifford algebra states,
	\[
	(\gamma^\mu p_\mu)(\gamma^\nu p_\nu) = p^2,
	\qquad
	p^2 = p_0^2 - p_1^2 - p_2^2 .
	\]
	
	We have to solve
	\[
	(-\,\gamma^\mu p_\mu)\,D(p) = -i .
	\]
	
	so multiply both sides by \(-\,\gamma^\mu p_\mu\):
	\[
	p^2\,D(p) = \,i\,\gamma^\mu p_\mu .
	\]
	
	Therefore,
	\[
	\boxed{
		D(p) = \,i\,\frac{\gamma^\mu p_\mu}{p^2}
	}
	\]
	
	\subsection*{Fourier transform identities}
	
	
	In \(d=3\) 
	\begin{equation}
		\int \frac{d^3 p}{(2\pi)^3}\,\frac{e^{ip\cdot x}}{p^2}
		= \frac{1}{4\pi |x|},
		\qquad x\neq 0 .
	\end{equation}
	
	
	Using
	\begin{equation}
		p_\mu e^{ip\cdot x} = -i\,\partial_\mu e^{ip\cdot x},
	\end{equation}
	we obtain
	\begin{equation}
		\int \frac{d^3 p}{(2\pi)^3}\,e^{ip\cdot x}\,
		\frac{p_\mu}{p^2}
		= -i\,\partial_\mu
		\left(\frac{1}{4\pi |x|}\right).
	\end{equation}
	
	Evaluating the derivative 
	\begin{equation}
		\partial_\mu \frac{1}{|x|}
		= -\frac{x_\mu}{|x|^3},
	\end{equation}
	we find, for \(x\neq 0\),
	\begin{equation}
		\boxed{
			\int \frac{d^3 p}{(2\pi)^3}\,e^{ip\cdot x}\,
			\frac{p_\mu}{p^2}
			= i\frac{x_\mu}{4\pi |x|^3}.
		}
	\end{equation}
	
	\section*{Tadpole integral evaluation }
	We evaluate the integral using a hard momentum cutoff.
	Consider
	\begin{equation}
		I_{\Lambda}
		=
		i \int_{|p|<\Lambda}
		\frac{d^3 p}{(2\pi)^3}
		\frac{1}{p^2 - m^2 + i\epsilon}.
	\end{equation}
	
	Under the Wick rotation $p^0 \to i p_E^0$, one has
	\begin{equation}
		p^2 - m^2 + i\epsilon \;\to\; - (p_E^2 + m^2),
		\qquad
		i\, d^3 p \;\to\; - d^3 p_E.
	\end{equation}
	The two minus signs cancel, yielding
	\begin{equation}
		I_{\Lambda}
		=
		\int_{|p_E|<\Lambda}
		\frac{d^3 p_E}{(2\pi)^3}
		\frac{1}{p_E^2 + m^2}.
	\end{equation}
	
	In three Euclidean dimensions,
	\begin{equation}
		d^3 p_E = 4\pi\, p^2\, dp,
	\end{equation}
	so the integral becomes
	\begin{equation}
		I_{\Lambda}
		=
		\frac{1}{(2\pi)^3}
		\int_0^\Lambda
		4\pi\,\frac{p^2\,dp}{p^2 + m^2}
		=
		\frac{1}{2\pi^2}
		\int_0^\Lambda
		\frac{p^2\,dp}{p^2 + m^2}.
	\end{equation}
	
	We rewrite the integrand as
	\begin{equation}
		\frac{p^2}{p^2 + m^2}
		=
		1 - \frac{m^2}{p^2 + m^2}.
	\end{equation}
	Therefore,
	\begin{align}
		\int_0^\Lambda \frac{p^2\,dp}{p^2 + m^2}
		&=
		\int_0^\Lambda dp
		-
		m^2 \int_0^\Lambda \frac{dp}{p^2 + m^2} \\
		&=
		\Lambda
		-
		m \int_0^{\Lambda/m} \frac{dp'}{p'^2 + 1} \\
		&=
		\Lambda
		-
		\frac{m\pi}{2} +	\mathcal{O}\!\left(\frac{m^3}{\Lambda^3}\right).  \\
	\end{align}

	or equivalently,
	\begin{equation}
		I_{\Lambda}
		=
		\frac{\Lambda}{2\pi^2}
		-
		\frac{m}{4\pi}
		+
		\mathcal{O}\!\left(\frac{m^2}{\Lambda}\right).
	\end{equation}

\end{document}